\documentclass[12pt]{elsart}

\usepackage{amsfonts}
\usepackage{xspace}

\usepackage{amsmath}

\usepackage{psfig,color,graphicx}

\newcommand{\smallpf}{{\mbox{\footnotesize pf}} }

\begin{document}

\begin{frontmatter}
\centerline{Submitted to Physica D, August 2003\hfill}

\title{Boundary effects and the onset of Taylor vortices}

\author{A.M.~Rucklidge}
\address{Department of Applied Mathematics,
University of Leeds, Leeds LS2 9JT, UK}
\and
\author{A.R.~Champneys}
\address{Department of Engineering Mathematics, University of Bristol,
Bristol BS8 1TR, UK}

\begin{abstract}
It is well established that the onset of spatially periodic vortex states in
the Taylor--Couette flow between rotating cylinders occurs at the value of
Reynold's number predicted by local bifurcation theory. However, the symmetry
breaking induced by the top and bottom plates means that the true situation
should be a disconnected pitchfork. Indeed, experiments have shown that the
fold of the disconnected branch can occur at more than double the Reynold's
number of onset.  This leads to an apparent contradiction: why should Taylor
vortices set in so sharply at the value Reynold's number predicted by the
symmetric theory, given such large symmetry-breaking effects caused by the
boundary conditions?  This paper offers a generic explanation. The details are
worked out using a Swift--Hohenberg pattern formation model that shares the
same qualitative features as the Taylor--Couette flow. Onset occurs via a wall
mode whose exponential tail penetrates further into the bulk of the domain as
the driving parameter increases. In a large domain of length~$L$, we show that
the wall mode creates significant amplitude in the centre at parameter values
that are $O(L^{-2})$ away from the value of onset in the problem with ideal
boundary conditions.  We explain this as being due to a Hamiltonian Hopf
bifurcation in space, which occurs at the same parameter value as the pitchfork
bifurcation of the temporal dynamics. The disconnected anomalous branch remains
$O(1)$ away from the onset parameter since it does not arise as a bifurcation
from the wall mode.
 \end{abstract}


\end{frontmatter}

\section{Introduction}

\begin{figure}
 \mbox{\psfig{file=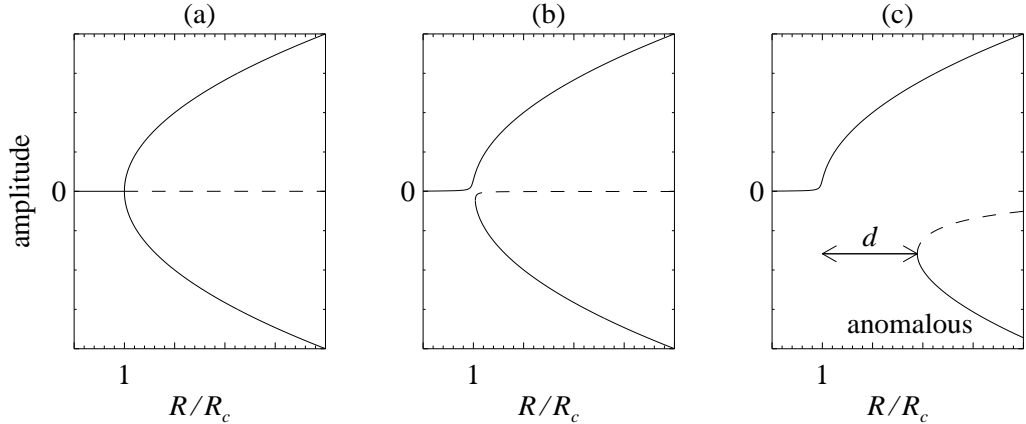,width=\hsize}}
 \caption{Sketches of the amplitude of Taylor vortices measured in the centre
of the apparatus as a function of Reynold's number~$R$, under various
assumptions.
 (a)~With ideal (reflecting or periodic) boundary conditions, there is a sharp
transition to Taylor vortices at a pitchfork bifurcation at $R=R_c$.
 (b)~With the assumption of a weakly broken pitchfork bifurcation, there is
still a relatively sharp onset of Taylor vortices close to $R=R_c$
(after~\cite{refB127a});
 (c)~The experiments of Benjamin and Mullin~\cite{refB128} suggest that the
upper half is still a weakly broken pitchfork, whereas the lower half is a
strongly broken pitchfork, with anomalous modes only appearing at Reynold's
numbers at least twice the critical value. Solid (dashed) lines indicate stable
(unstable) solutions.}
 \label{fig:brokenpitchfork}
 \end{figure}

The Taylor--Couette experiment provided one of the first quantitative
verifications of the correctness of the Navier--Stokes partial differential
equations (PDEs) describing the dynamics of fluid flows. The experiment, in its
simplest form, consists of a pair of concentric cylinders with a fluid-filled
gap in between; as the inner cylinder is rotated, a shearing flow (the Couette
flow) is established between the cylinders, and this becomes unstable to
axisymmetric vortices (Taylor vortices) at a critical value of the rotation
rate (as measured by a dimensionless Reynold's number~$R$).
See~\cite{refD55,Ko:93} for reviews. One notable achievement of Taylor's
work~\cite{refT61} in 1923 was the theoretical prediction and the experimental
measurement of the critical Reynold's number~$R_c$ for the onset of vortices,
with remarkably good agreement between the two. In doing the stability
calculation, Taylor assumed that the vortices would be periodic in the
direction along the axis, and neglected the effects of the top and the bottom
plates of the experiment. With this assumption, the bifurcation leading to
Taylor vortices is a pitchfork (figure~\ref{fig:brokenpitchfork}a), and the
characteristic sharp transition of this bifurcation, with the strength of the
vortices going as the square root of the degree of supercriticality $R-R_c$,
has been confirmed experimentally~\cite{refG97}. The symmetry that is broken in
this pitchfork is a translation symmetry along the axes of the cylinders.

\begin{figure}
 \begin{center}
 \mbox{\psfig{file=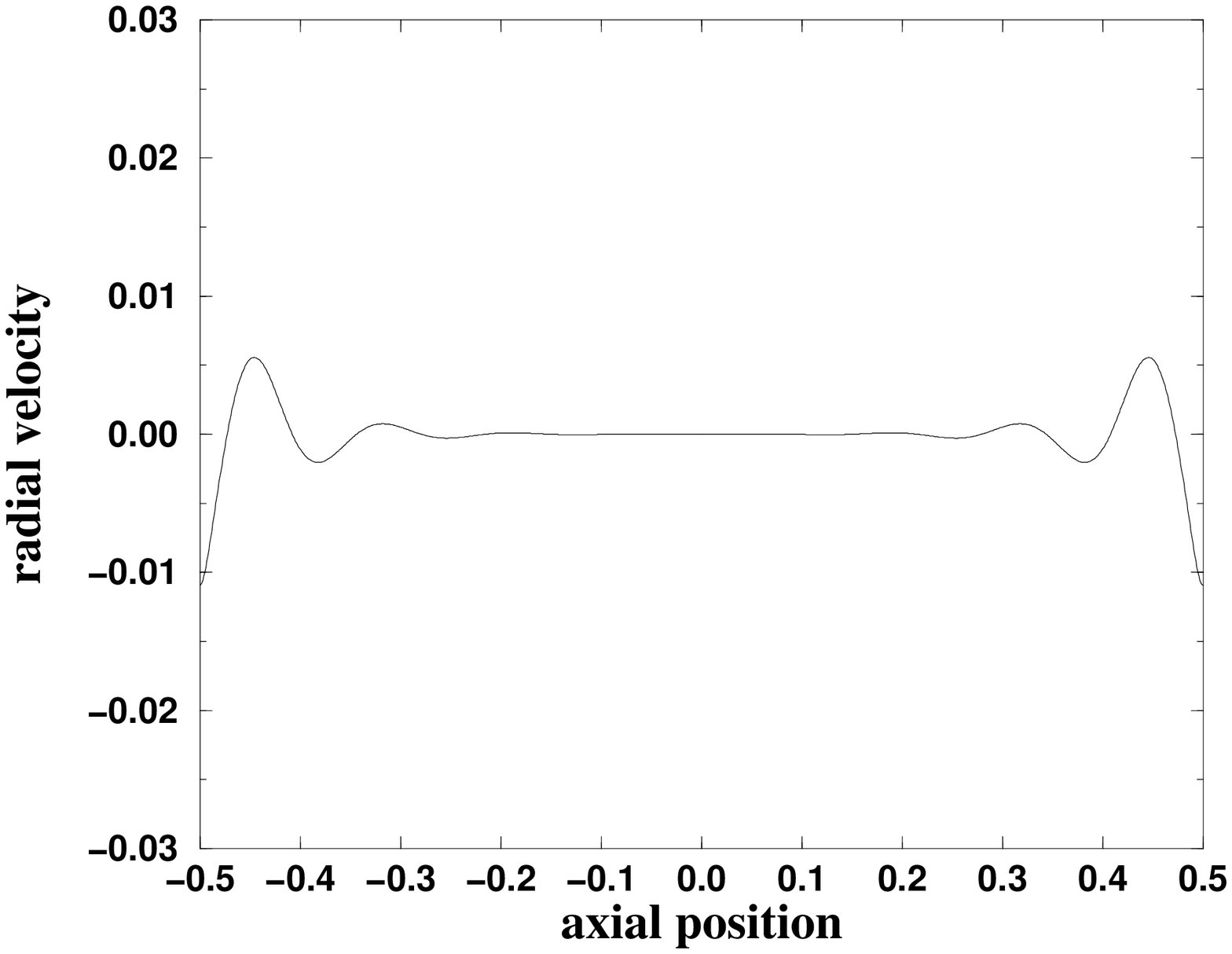,width=0.45\hsize}}
 \mbox{\psfig{file=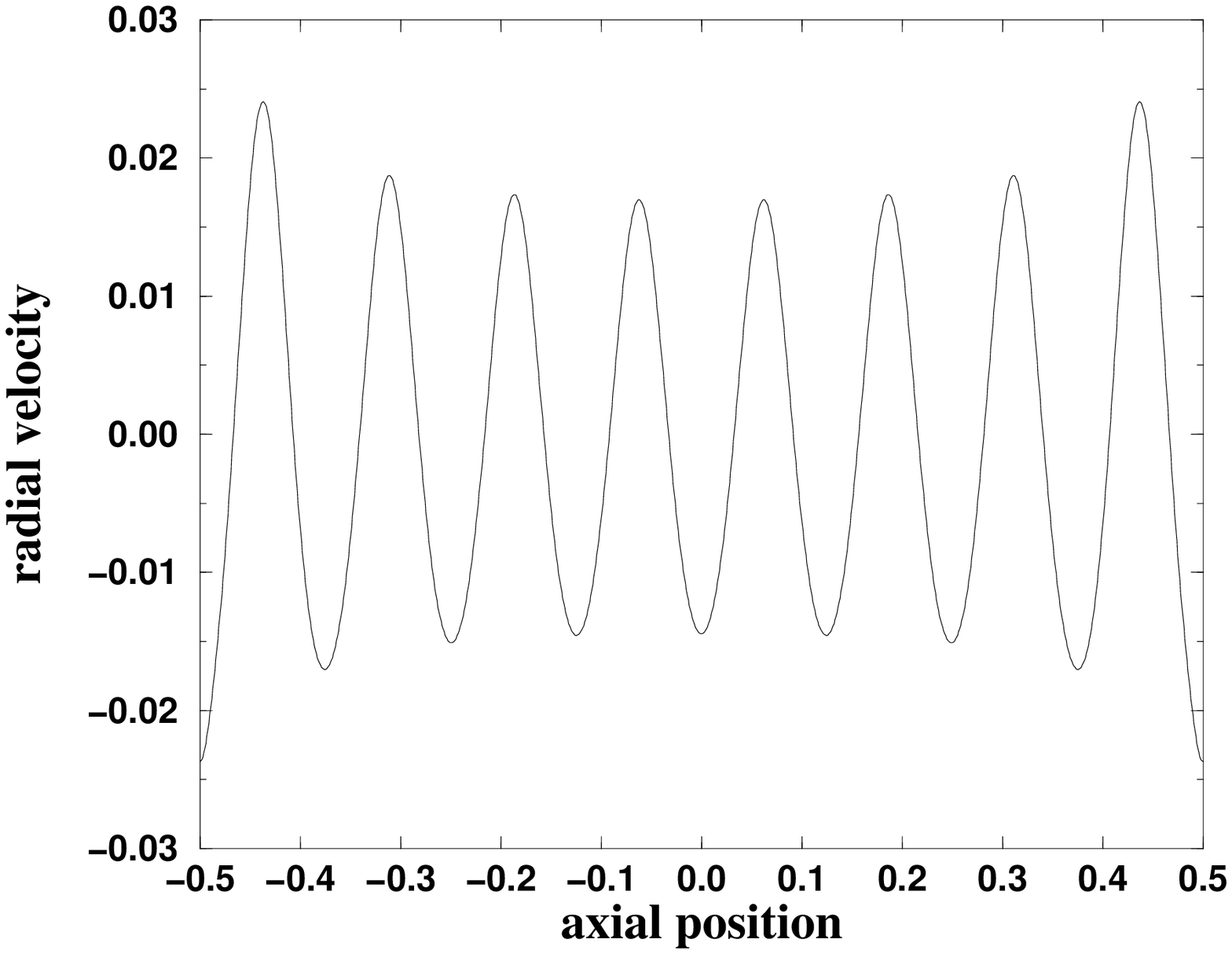,width=0.45\hsize}}
 \end{center}
\caption{Finite-element results reproduced with permission from~\cite{refA70}
showing two radial velocity profiles below and above the critical value of~$R$
at which the onset of periodic vortices would occur with periodic boundary
conditions.
 (a)~$R/R_c=63.74/68.189=0.935$;
 (b)~$R/R_c=69.02/68.189=1.013$.}
 \label{fig:abshagen}
 \end{figure}

Subsequent theoretical developments explored the role of the top and bottom
plates in the experiment, which spoil the idealisation of spatial periodicity
in the direction parallel to the axis of rotation, and which break the
translation symmetry assumed in the original theoretical work. Ekman boundary
layers cause the fluid near the boundary to spiral preferentially inwards for
any non-zero rotation rate, and the Taylor vortices to develop first in the
boundary layer, moving smoothly into the bulk of the fluid as $R$
approaches~$R_c$ -- this has been observed in experiments~\cite{refC124} and in
calculations~\cite{refA71}. In the words of Benjamin~\cite{refB127a}, `no
precise critical value of~$R$ exists for the onset of cellular motion'.
Benjamin~\cite{refB127a} interpreted the formation of Taylor vortices as a
broken pitchfork (figure~\ref{fig:brokenpitchfork}b), with the end plates
driving a flow near the boundary for all non-zero~$R$, and this flow exciting a
cellular flow that penetrates the central region with increasing Reynold's
number -- see figure~\ref{fig:abshagen}.

The qualitative role of the end boundaries was explored further by
Schaeffer~\cite{refS111}, who introduced a homotopy parameter~$\tau$
($0\leq\tau\leq1$), with $\tau=0$ corresponding to ideal end boundary
conditions, where a state of pure Couette flow exists for all values of~$R$,
$\tau=1$ corresponding to physically realistic end boundary conditions, and
intermediate values of~$\tau$ interpolating between these two extremes. The
results for $\tau$~close to zero are indeed consistent with a weakly broken
pitchfork bifurcation (figure~\ref{fig:brokenpitchfork}b), and are in
qualitative agreement with experimental results of
Benjamin~\cite{refB127a,refB127b}: as the Reynold's number is increased slowly,
vortices grow smoothly, with the most rapid growth occurring for $R$~close to
the critical value. This is illustrated in figure~\ref{fig:abshagen}: note how
the radial velocity profiles are not zero for $R$ below~$R_c$; instead we see a
pair of `wall modes' connecting the non-parallel flow at the walls to Couette
flow in the main body of the cylinder. Similarly, the steady solution for $R$
greater than~$R_c$ is not a pure periodic solution but has modulation near the
two walls in order to satisfy the inhomogeneous boundary conditions.

However, by starting the experiment impulsively, it is possible to find another
branch of vortices ({\it anomalous modes}) that resemble ordinary vortices, but
have the opposite sign -- close to the ends, these anomalous vortices have an
outwards radial velocity, opposite to the normal vortices~\cite{refB127b}.
Other experiments and calculations indicate that anomalous modes may also have
stagnation regions or narrow counter-cells close to the top and bottom
boundaries~\cite{refB129,refC123}.

However, anomalous modes cannot be found close to the critical Reynold's
number, as would be suggested by figure~\ref{fig:brokenpitchfork}b. Instead,
$R$~needs to be at least twice its critical value (and the cylinder started
impulsively) before anomalous modes can be found~\cite{refB128}. Once they are
established, the anomalous modes persist as $R$ is decreased to a lower
stability bound, the exact value of which depend on experimental parameters
such as the gap width between the inner and outer cylinders, or the aspect
ratio~$L$, a dimensionless measure of the length of the column. The lower
existence boundary appears always to be at least twice the critical
value~$R_c$~\cite{refB128}, and can be much higher if the gap between the
cylinders is narrow~\cite{refC122}. Thus the experimental situation is depicted
in figure~\ref{fig:brokenpitchfork}c. Interestingly, the lower stability
boundary of the anomalous modes seems to be independent of the aspect ratio of
the apparatus (for large aspect ratios), and remains at an appreciable multiple
of the critical Reynold's number for ordinary modes even as $L$ becomes
large~\cite{refC122,refL66}. This emphasises the fact that the large (but
finite) aspect ratio limit is very different from the idealisation of
periodicity in the axial direction.

It is worth emphasising the two surprising and apparently contradictory aspects
that have emerged. First, the distance ($d$~in
figure~\ref{fig:brokenpitchfork}(c)) between the fold on the disconnected
(anomalous) branch is such that the Reynold's number at the fold is at least a
factor of two larger than the Reynold's number of the onset of ordinary Taylor
vortices. Making the cylinder longer (so that the boundary effects are moved
`towards infinity') does not make $d$ tend to zero, and many authors have
concluded that the onset of Taylor vortices can in no way be regarded as a
weakly broken pitchfork. Second, the onset of vortices is sharp when viewed in
terms of measures such as the radial velocity of the midpoint of the apparatus,
and it occurs at almost exactly the value of $R$ that is predicted for the
problem without end effects -- so the onset of Taylor vortices apparently can
be described as a weakly broken pitchfork bifurcation. This discrepancy would
not be expected for a generic unfolding of a pitchfork bifurcation.  The
purpose of the present paper is to explain this apparent contradiction.

The rest of the paper is outlined as follows. In section~2, we introduce the
Swift--Hohenberg equations as a model for the Taylor--Couette flow. We note
that the modification of pattern formation due to the presence of weak forcing
at lateral boundaries in Swift--Hohenberg equations has been addressed in the
work of Daniels and co-workers~\cite{refD56,DaHoSk:03} with application to
Rayleigh--B\'enard convection in mind. In contrast with their work, we include
an order one boundary condition that forces the flow strongly. This approach is 
also distinguished from that of~\cite{refP58}, which focused on a
Ginzberg--Landau equation for the envelope of the vortex amplitude.

Section~3 then includes an analysis of the linearised version of this model.
Despite being purely linear, it is found that mode shapes like those in
figure~\ref{fig:abshagen} emerge under static increase of the bifurcation
parameter through the value at which the symmetric problem bifurcates. We
explain this in terms of the relation between the temporal pitchfork
bifurcation and a spatial Hamiltonian Hopf bifurcation.

Section~4 goes on to consider a nonlinear bifurcation analysis. It transpires
that the unfolded pitchfork resembles figure~\ref{fig:brokenpitchfork}(c).
There are many anomalous branches that emerge from primary and secondary
symmetry breaking bifurcations in the symmetric problem. In the 
Swift--Hohenberg example studied in detail, the stable anomalous branch does
not always emerge from the primary pitchfork bifurcation when the
symmetry-breaking terms tend to zero. We explore this issue in some detail.
Finally, section~5 draws conclusions and discusses wider implications of the
results.

\section{Swift--Hohenberg model}

Rather than consider the nonlinear axisymmetric hydrodynamic partial
differential equations (PDEs) that describe the flow between two rotating
cylinders, we focus on the simpler Swift--Hohenberg~\cite{refS109} PDE, which
shares many of the same pattern-forming features. In fact,
Melbourne~\cite{refM99} has demonstrated that bifurcation problems of the
Taylor--Couette type (steady state bifurcations with nonzero critical
wave\-number in systems with Euclidean symmetry) reduce to equations of
Swift--Hohenberg form (though with more general nonlinear terms). The model
equation we use is:
 \begin{equation}
 U_t=\mu U - (1+2U_{xx}+U_{xxxx}) - U^3 - U\,U_x,
 \label{eq:SH}
 \end{equation}
where the subscripts denote partial derivatives with respect to space~$x$ and
time~$t$. The dependent variable $U(x,t)\in\Rset$ is defined on
$x\in[-L/2,L/2]$, where $L$ represents the length of the Taylor column. The
parameter~$\mu$ represents the forcing~$R-R_c$. The usual form of the
Swift--Hohenberg equation has only a cubic nonlinearity, but we include a
quadratic term to ensure there is no $U\rightarrow-U$ symmetry
(see, e.g., \cite[eq.~(7.21)]{Ko:93}. This term also
makes the model equation non-variational and so allows unsteady behaviour as 
an asymptotic state (though we focus entirely on steady states).

In order to relate the order parameter~$U$ to the fluid flow, we interpret~$U$
as a stream function, and so $U_x$ represents the radial velocity in the
column. The effect of the non-slip boundary conditions on the top and bottom 
plates is to induce an inwards flow 
near the boundaries, although the radial velocity is zero on the boundaries 
themselves. We model this 
strong forcing at the end walls with the inhomogeneous boundary conditions:
 \begin{equation}
 U(-L/2)=U(L/2)=0,\qquad 
 U_x(-L/2)=U_x(L/2)=-1.
 \label{eq:BCreal}
 \end{equation}
The only symmetry in the problem is the reflection in the equatorial mid-plane
of the apparatus: 
\begin{equation}
(U,x)\rightarrow(-U,-x).
\label{eq:symmetry}
\end{equation}

We also consider idealised reflecting boundary conditions:
 \begin{equation}
 U(-L/2)=U(L/2)=0,\qquad 
 U_{xx}(-L/2)=U_{xx}(L/2)=0, 
 \label{eq:BCreflecting}
 \end{equation}
which have an additional hidden symmetry~\cite{refG98}: the problem can be
extended by reflection onto the domain $x\in[-L,L]$ with periodic boundary
conditions and so acquires a continuous translation symmetry. It is this
translation symmetry that is broken in the pitchfork bifurcation in the
idealised version of this problem (figure~\ref{fig:brokenpitchfork}a). This
symmetry is strongly broken by the inhomogeneous boundary
conditions~(\ref{eq:BCreal}).

Experiments find states where the mid-plane reflection symmetry is preserved,
so we focus on this case by using reflecting boundary conditions at $x=0$. We
are also interested in examining the transition from idealised to realistic
boundary conditions, and so we use:
 \begin{equation}
 U(0)=U_{xx}(0)=U(L/2)=
 \tau\left(U_x(L/2)+1\right) + (1-\tau)U_{xx}(L/2)=0.
 \label{eq:BCboth}
 \end{equation}
Here $\tau$ ($0\leq\tau\leq1$) is a homotopy parameter~\cite{refS111}, such
that $\tau=0$ corresponds to the ideal problem (with hidden translation
symmetry) and $\tau=1$ to the realistic non-slip boundary conditions.

\section{Linear analysis: pitchfork and Hamiltonian Hopf}

Solutions of the linearised problem can be written in terms of exponentials in
time and space, in the form:
 \begin{equation}
 U(x,t)=e^{st + (\sigma+ik)x},
 \end{equation}
where $s$ is the temporal growth rate, $k$ is a spatial wave\-number, and
$\sigma$ is a spatial growth rate. In order to satisfy the linearised model
equation, $s$, $\sigma$ and $k$ must satisfy
 \begin{equation}
 0=k\sigma(k^2-\sigma^2-1),\qquad
 s=\mu-(k^2-\sigma^2-1)^2+4\sigma^2k^2,
 \end{equation} 
which can be rearranged to give three possibilities:
 \begin{eqnarray}
 A:& \quad k=0,            \qquad &s=\mu-(1+\sigma^2)^2,\\
 B:& \quad \sigma=0,       \qquad &s=\mu-(1-k^2)^2,\\
 C:& \quad k^2=\sigma^2+1, \qquad &s=\mu+4\sigma^2(1+\sigma^2).
 \end{eqnarray} 
Since the linearisation of the PDE~(\ref{eq:SH}) is first order in time and
fourth order in space, there is a unique temporal growth rate~$s$ and a total
of four complex spatial growth rates, which are roots of the equations above,
namely in case~A: $\pm\sigma_1$ and $\pm\sigma_2$; in case~B: $\pm ik_1$ and
$\pm ik_2$; in case~C: $\pm\sigma\pm ik$. Corresponding to these, the linear
solutions are of the form:
 \begin{eqnarray}
 A:& \quad U(x,t)=e^{st}\left(A\sinh(\sigma_1x)+B\sinh(\sigma_2x)\right),\\
 B:& \quad U(x,t)=e^{st}\left(A\sin(k_1x)+B\sin(k_2x)\right),\\
 C:& \quad U(x,t)=e^{st}\left(A\cos(kx)\sinh(\sigma x) + 
                              B\sin(kx)\cosh(\sigma x)\right),
 \end{eqnarray} 
where $A$ and $B$ are constants that will be determined by the boundary
conditions at $x=L/2$. The odd boundary conditions at $x=0$ have already
been enforced by the choice of trigonometric functions.

The next stage of the calculation depends on whether ideal ($\tau=0$) or
realistic ($\tau=1$) boundary conditions are being used. With ideal boundary
conditions, $U=0$ is always a solution of~(\ref{eq:SH}), and bifurcations from
this state occur when there are marginally stable ($s=0$) linear solutions.
Setting $s=0$ and $U(L/2)=U_{xx}(L/2)=0$ results in no solution in cases~A
and~C, and an eigenvalue problem in case~B, where $k$ can take on discrete
values: $k=2\pi n/L$, where $n$ is the number of vortices in the half-domain.
There are thus pitchfork bifurcations at
 \begin{equation}
 \mu_{\smallpf}=\left(1-\left(\frac{2\pi n}{L}\right)^2\right)^2,
 \label{eq:mupitchfork}
 \end{equation}
where $n$ is an integer. In particular, there is a pitchfork bifurcation at
$\mu=0$ whenever the domain is chosen to fit an exact number of vortices:
$n=L/2\pi$. Note that the condition for the onset of the vortices is $s=0$ and
$\sigma=0$, with a purely imaginary spatial wave\-number $\pm ik$ corresponding
to a spatially periodic pattern. On either side of this bifurcation point, the
temporal growth rate~$s$, indicating the stability of the Couette flow, changes
from negative to positive at the pitchfork bifurcation.

\begin{figure}
\begin{center}
\mbox{\psfig{file=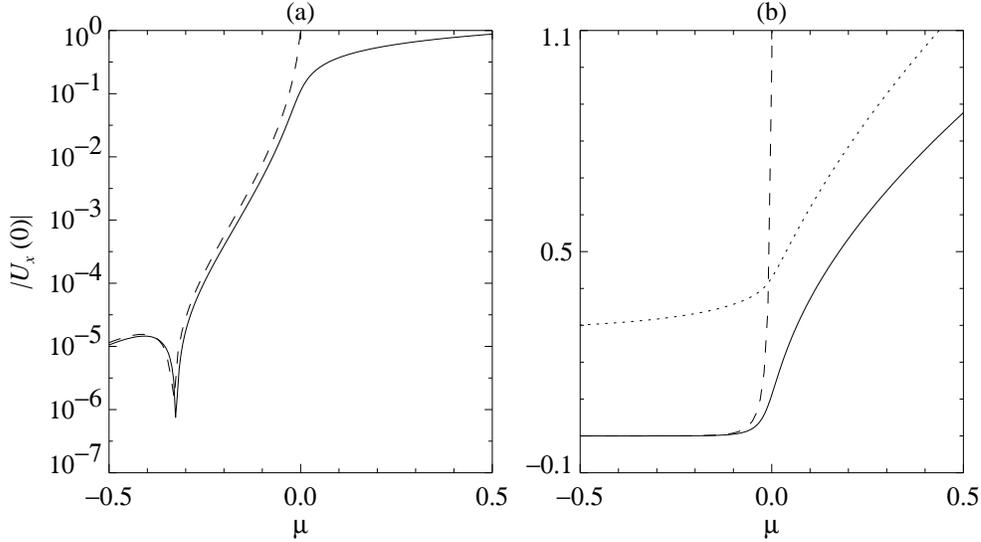,width=0.9\hsize}}
\end{center} \caption{Dependence of the radial velocity at the midpoint
($U_x(0)$) as a function of bifurcation parameter~$\mu$ for the nonlinear
(solid) and linearised (dashed) PDE~(\ref{eq:SH}), with $L=22\pi$ and realistic
boundary conditions ($\tau=1$). Also shown in (b) as a dotted line is an
integrated average of the solution and its first three derivatives, showing how
other measures of amplitude are not so sharp as the radial velocity at the
midpoint.}
 \label{fig:lin_nonlin_bifn}
 \end{figure}

On the other hand, with inhomogeneous boundary conditions ($\tau=1$), there are
nonzero steady ($s=0$) linear solutions for all values of~$\mu$. Case~A can
only arise when $\mu\geq1$ (and in fact only a single solution is possible); in
case~B, there are two possible solutions for $0\geq\mu\geq1$ and one for
$\mu>1$, and case~C is possible only when~$\mu\leq0$. The solutions are thus,
for $\mu<0$: $U(x)=A\cos(kx)\sinh(\sigma x) + B\sin(kx)\cosh(\sigma x)$, where
$\sigma>0$ and $k>0$ are determined from $\mu$ by $4\sigma^4+4\sigma^2+\mu=0$
and $k^2=1+\sigma^2$; and for $0<\mu<1$: $U(x)=A\sin(k_1x)+B\sin(k_2x)$, where
$k_1>0$, $k_2>0$ and $k_1^2=1+\sqrt{\mu}$, $k_2^2=1-\sqrt{\mu}$. In these two
expressions for $U(x)$, the constants $A$ and $B$ are determined by a pair of
linear equations from the boundary conditions $U(L/2)=0$ and $U_x(L/2)=-1$. The
dependence of the solution, as measured by the radial velocity at the midpoint
($U_x(0)=A\sigma + Bk$ for $\mu<0$, and $U_x(0)=Ak_1+Bk_2$ for $0<\mu<1$) is
shown in figure~\ref{fig:lin_nonlin_bifn} (dashed line). 

The amplitude of the linear solution goes to infinity for $\mu$~about $0.00826$
(with~$L=22\pi$). Since there is no sharp onset with realistic boundaries, one
cannot define a precise value of~$\mu$ at which pattern will be first observed
in a domain of finite length, but the value of $\mu=\mu_{\infty}$ for which the
linear solution goes to infinity is a suitable proxy. This is defined 
implicitly by the condition
 \begin{equation}
 (k_1+k_2)\sin\left((k_1-k_2)\frac{L}{2}\right)= 
 (k_1-k_2)\sin\left((k_1+k_2)\frac{L}{2}\right),
 \label{eq:muinfinity}
 \end{equation}
where $k_{1,2}$ are defined in terms of~$\mu$ above. 
When $L$ is large, the smallest positive solution $\mu_{\infty}$ occurs
when $(k_1-k_2)L/2 = \pi$. This yields
 \begin{equation}
 \mu_{\infty} = \frac{4\pi^2}{L^2} + O(L^{-3}).
 \end{equation}
In other words, the divergence of the linear solution occurs for $\mu$~closer
to zero as $L$ increases. Moreover, we have that for large~$L$ the first
pitchfork bifurcation occurs at
 \begin{equation}
 \mu_{\smallpf} = \dfrac{4\delta^2}{L^2} + O(L^{-3}), \quad 
 \mbox{where} \quad \delta = L- 2 \pi \left [ L/(2\pi) \right], 
 \end{equation}
so $|\delta|<2\pi$. Hence $\mu_{\infty}$, $\mu_{\smallpf}$ and the difference
between them are all of the same order:~$L^{-2}$.

Also shown in figure~\ref{fig:lin_nonlin_bifn} are nonlinear solutions
of~(\ref{eq:SH}). The linear and nonlinear solutions are close to each other
for $\mu$~negative: the discrepancy at low amplitude arises because the
linearised solution is not small close to the boundaries. 

\begin{figure}
\begin{center}
\mbox{\psfig{file=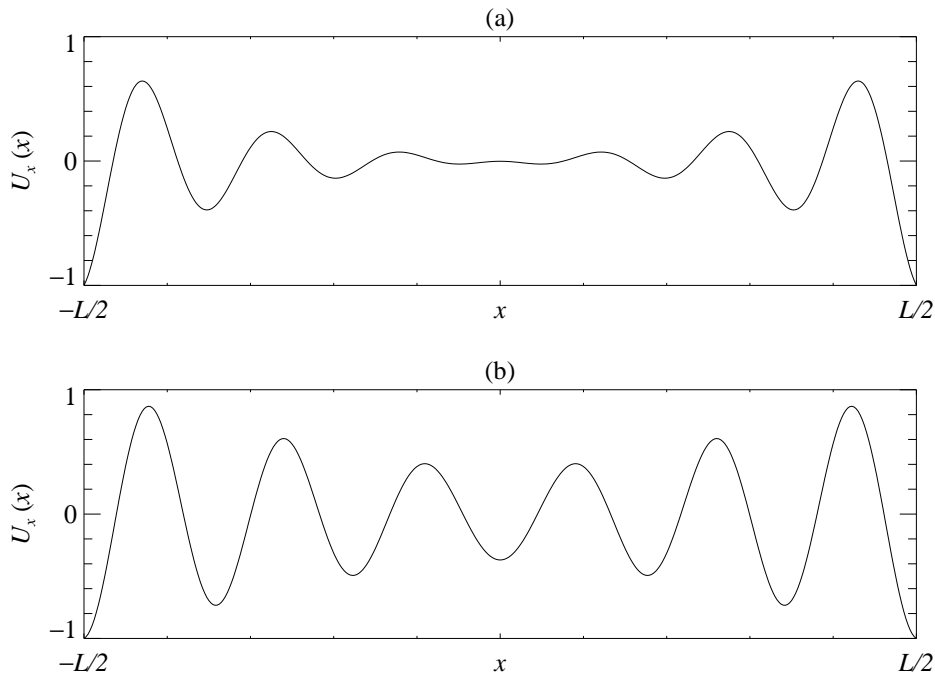,width=0.9\hsize}}
\end{center}
\caption{Linear solutions of~(\ref{eq:SH}), with $L=40$ and~$\tau=1$, with
(a)~$\mu=-0.1$ and (b)~$\mu=0$, showing the growth of the solution in from the
edges ($x=\pm L/2$). Compare with figure~\ref{fig:abshagen}.}
 \label{fig:lin_soln}
 \end{figure}

\begin{figure}
\begin{center}
\mbox{\psfig{file=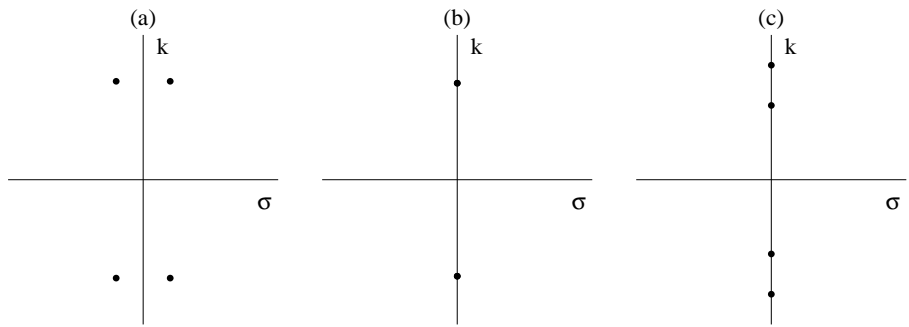,width=0.9\hsize}}
\end{center}
\caption{Hamiltonian Hopf bifurcation:
 (a)~$\mu<0$, the spatial eigenvalues are $\pm\sigma\pm ik$;
 (b)~$\mu=0$, the eigenvalues $\pm i$;
 (c)~$\mu>0$, the eigenvalues are $\pm ik_1$, $\pm ik_2$.}
 \label{fig:HamiltonianHopf}
 \end{figure}

Linear solutions of the PDE~(\ref{eq:SH}) are shown in
figure~\ref{fig:lin_soln} for two values of~$\mu$. As $\mu$ is increased from
negative values to $\mu=0$, the exponentially decaying linear solution extends
further into the bulk of the fluid. In an arbitrarily long cylinder, the radial
velocity at the midpoint of the apparatus would remain almost zero until the
spatial decay rate (as measured by~$\sigma$) became zero. Therefore, the
condition for onset of steady vortices, as measured at the midpoint of the
apparatus, is $\sigma=0$ and $s=0$ -- which is the same condition as for the
onset of vortices with the idealised boundary conditions. In this case,
however, on either side of onset, it is the spatial eigenvalues $\pm\sigma\pm
ik$ and $\pm ik_1$, $\pm ik_2$ that change in nature, at a Hamiltonian Hopf
bifurcation \cite{vdeMe:85} (see figure~\ref{fig:HamiltonianHopf}). 

More accurately, we should describe the spatial bifurcation as a reversible 1:1
resonance, since the ODE obtained by setting $U_t=0$ in (\ref{eq:SH}) does not
conserve a first integral and so cannot correspond to a Hamiltonian system, but
is nevertheless reversible in the sense analysed by Iooss and Peroueme
\cite{IoPe:93} owing to the symmetry (\ref{eq:symmetry}). Here the spatial
bifurcation is supercritical and for $\mu>0$ there exist spatially periodic
solutions the maximum amplitude of which grows as the square root of~$\mu$ for
the nonlinear problem. The implications of this bifurcation are discussed in
more detail in Section~5.

This explanation carries over from the Swift--Hohenberg model to the
Taylor--Couette problem and more general pattern forming situations
with strong forcing at the boundaries. If one assumes periodic
boundary conditions, with pure imaginary spatial wave\-numbers, the
criterion for onset in a general pattern forming problem is that the
temporal growth rate is zero. If, one the other hand, one takes the
end walls into account but the domain is very large, and the pattern
is measured only far away from the boundaries, then the steady
inhomogeneous solution will penetrate into the bulk of the fluid and
reach the centre when the real part of the spatial wave\-number is
zero. Thus the two perspectives will yield the same condition for the
onset of pattern formation: $\mu=0$ in the case of the model PDE, or
$R=R_c$ in the case of Taylor--Couette flow. This explains the sharp
transition seen in large domain Taylor--Couette experiments at the
Reynold's number predicted using idealised boundary conditions, even
though the boundaries are forcing the flow strongly.

\section{Nonlinear steady-state bifurcation analysis}

The remaining issues to be addressed are the effect of the length 
of the domain on the nonlinear solutions, and the location of the
saddle-node bifurcation on the anomalous branch
(figure~\ref{fig:brokenpitchfork}c).

\begin{figure}
\begin{center}
\mbox{\psfig{file=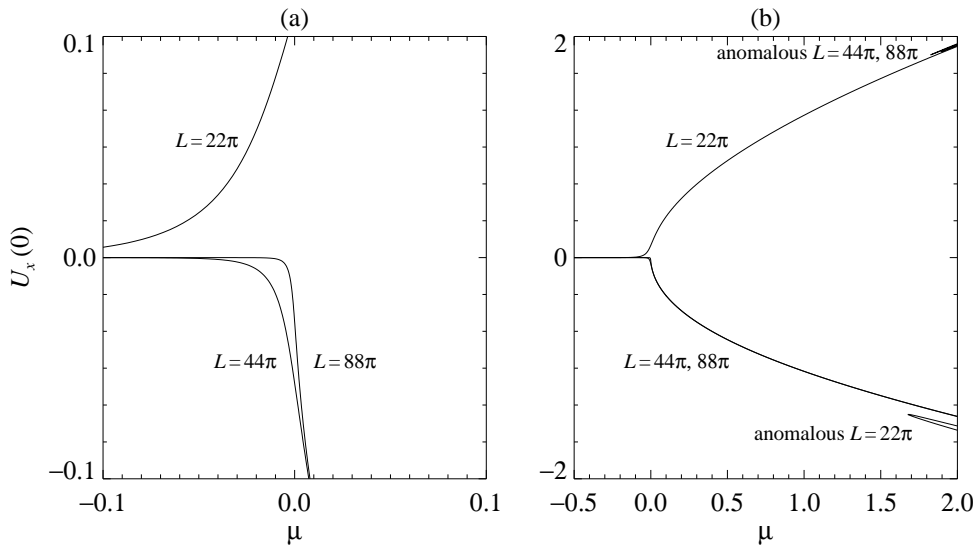,width=0.9\hsize}}
\end{center}
\caption{As the length of the domain increases, the pattern, as
measured in the centre, sets in more sharply: (a)~detail near $\mu=0$,
for $L=22\pi$, $44\pi$ and $88\pi$; (b)~the larger picture, showing
the anomalous modes.}
\label{fig:bifn_length} 
\end{figure}

\begin{figure}
\begin{center}
\mbox{\psfig{file=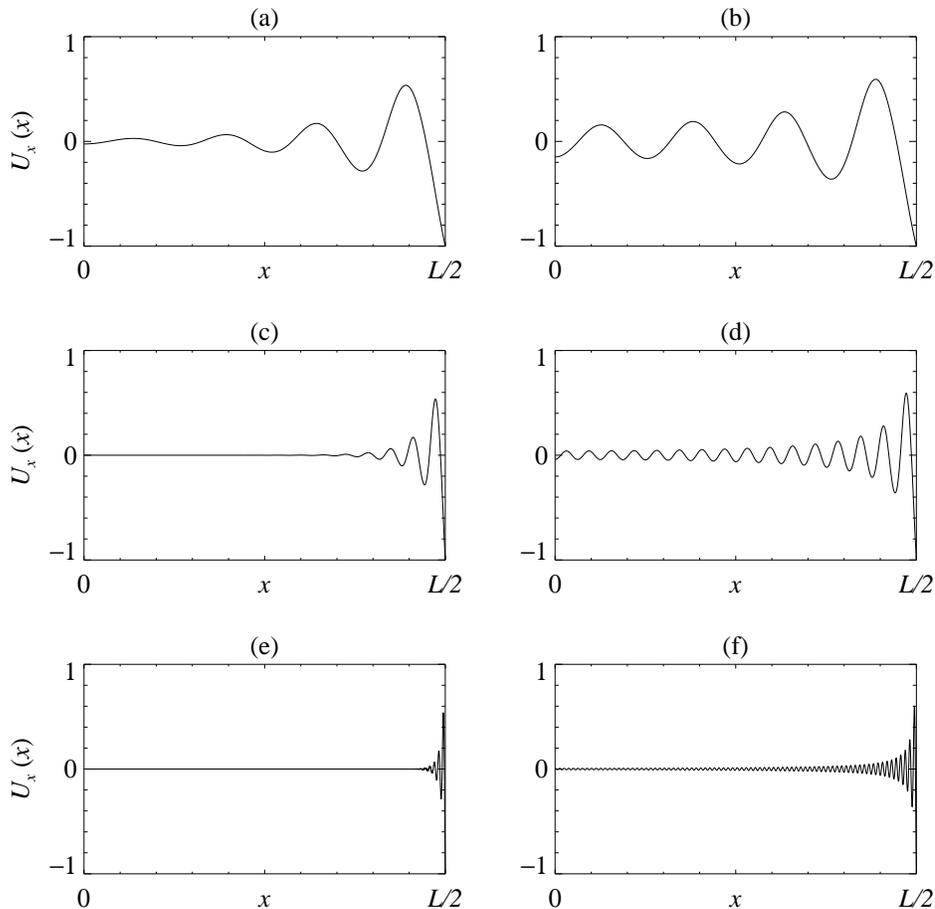,width=0.9\hsize}}
\end{center}
\caption{As $\mu$ increases through zero, the wall mode penetrates
further into the bulk, but the amplitude at the centre of the domain remains 
small for larger values of~$L$.
 (a,b)~$L=50$, $\mu=-0.1$,~$0$;
 (c,d)~$L=200$, $\mu=-0.1$,~$0$;
 (e,f)~$L=1000$, $\mu=-0.1$,~$0$.
 These are nonlinear solutions, but linear solutions look very similar.}
\label{fig:nonlin_length}
\end{figure}

We focus on the steady state problem given by setting $U_t=0$
in~(\ref{eq:SH}).
 \begin{equation}
 U_{xxxx} +2 U_{xx} + (1-\mu) U + U^3 + U\,U_x =0,
 \label{steady}
 \end{equation} 
subject to boundary conditions~(\ref{eq:BCboth}). Nonlinear solutions are
computed using AUTO \cite{Auto} as a boundary value solver. The effect of the
size of the domain is illustrated in figures~\ref{fig:bifn_length}
and~\ref{fig:nonlin_length}. As the length~$L$ increases, the amplitude, as
measured in the centre, sets in more sharply as $\mu$~is increased through
zero, and the curve resembles half a pitchfork as $L\rightarrow\infty$. With
very large values of~$L$ (figure~\ref{fig:nonlin_length}), the exponential
decay into the bulk ensures that the inhomogeneous pattern has very small
amplitude for~$\mu\leq0$. This is consistent with our understanding from the
linear theory. The anomalous mode branches for $L=22\pi$, $44\pi$ and $88\pi$
are also shown in figure~\ref{fig:bifn_length}. Note how the anomalous modes do
not approach $\mu=0$ for larger values of~$L$, even though the 
symmetry-breaking effects are being pushed further away.

With a fixed value of~$L$, only one sign of $U_x(0)$ is possible near the
transition at $\mu=0$, but which one is observed depends whether the number of
vortices between $x=0$ and $x=L/2$ is even or odd.
 
\begin{figure}
\begin{center}
\mbox{\psfig{file=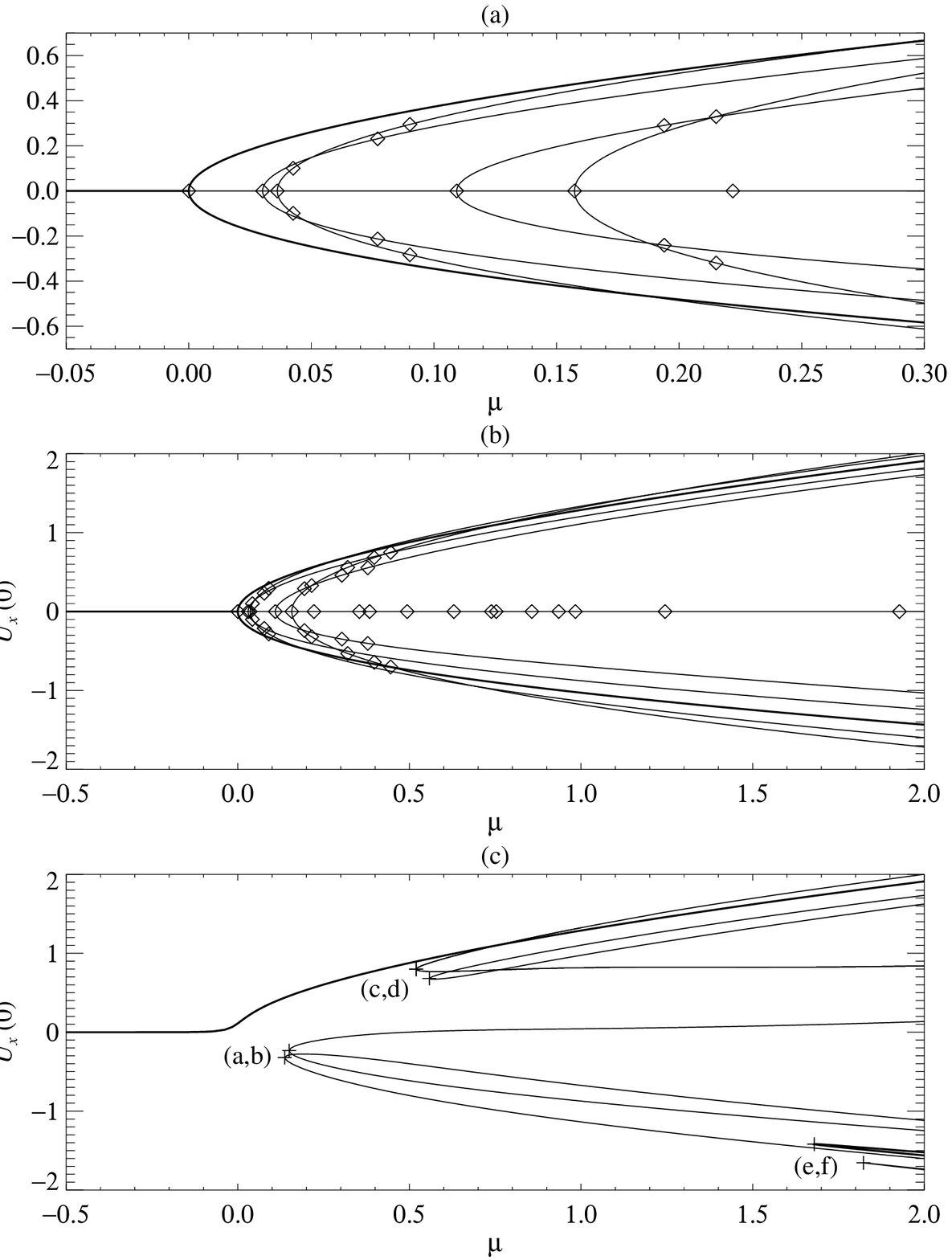,width=0.9\hsize}}
\end{center}
\caption{(a,b)~The idealised bifurcation diagram with $\tau=0$ and
$L=22\pi$: $U_x(0)$ as a function of bifurcation parameter~$\mu$. Diamonds
represent pitchfork bifurcation points. Only the first five bifurcating
branches from the trivial solution are depicted. In increasing order of~$\mu$, 
these have 11, 10, 12, 9 and 13~pairs of 
vortices in the full domain. (c)~Bifurcation diagram with $\tau=1$, showing the 
smooth onset of the 11~vortex solution, and several disconnected branches. 
The plus signs indicate solutions that are depicted in 
figure~\ref{fig:lp_examples}(a--f), at saddle-node 
bifurcation points.
The thick lines represent branches that are known to be stable.}
\label{fig:baseline} 
\end{figure}

\begin{figure}
\begin{center}
\mbox{\psfig{file=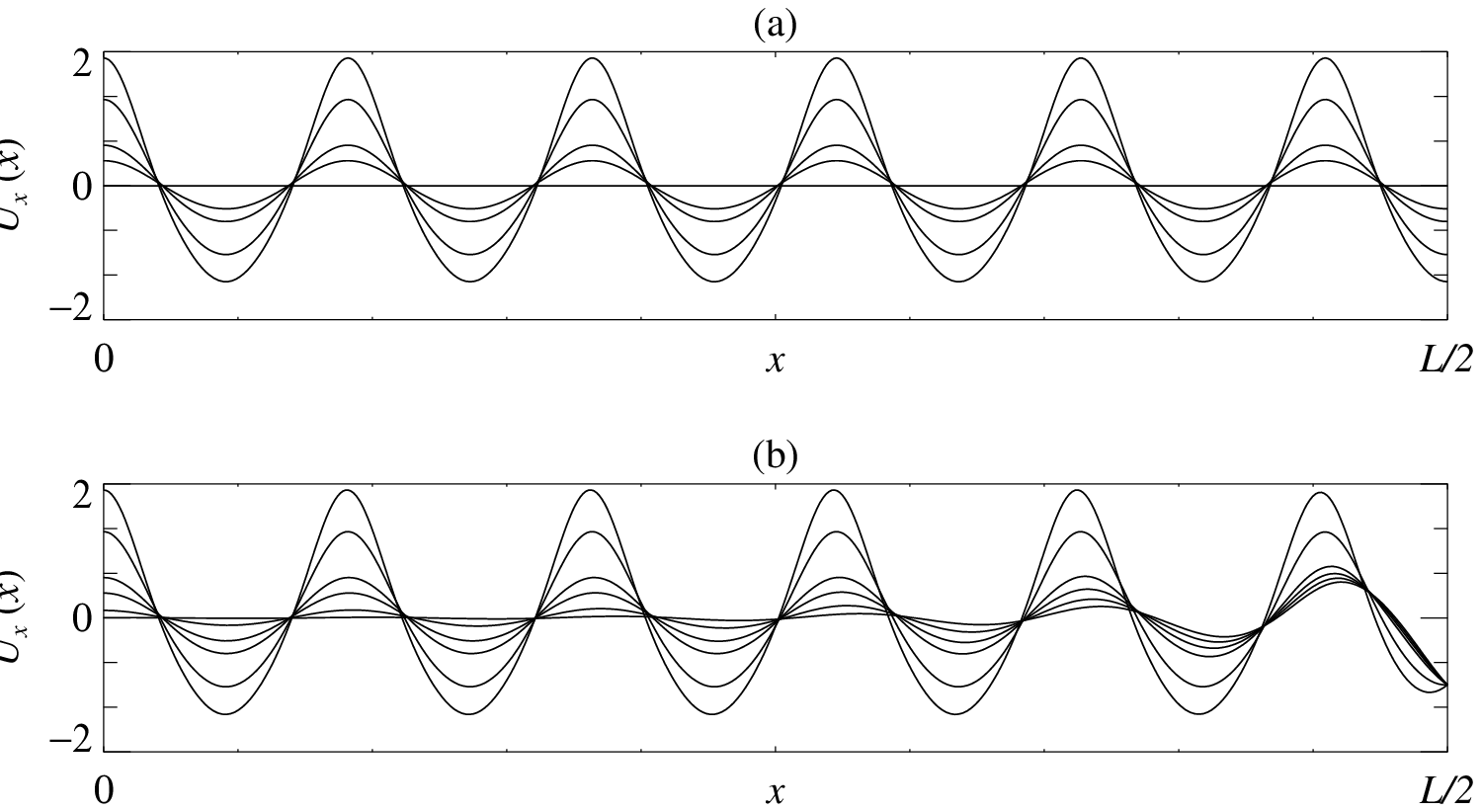,width=0.9\hsize}}
\end{center}
\caption{Examples of $n=11$~vortex solutions with (a) ideal ($\tau=0$) and
(b)~realistic ($\tau=1$) boundary conditions, with $\mu=2$ (largest amplitude), 
$\mu=1$, $\mu=0.25$, $\mu=0.1$, $\mu=0$, $\mu=-0.1$.}
\label{fig:ideal_real_examples}
\end{figure}

Next, we consider the connection between pitchfork bifurcations (in the case of
ideal boundary conditions) and saddle-node bifurcations (in the case of
realistic boundary conditions). With ideal boundary conditions, pitchforks
occur both from the trivial solution and as secondary bifurcations from the
various primary branches. We concentrate on the case $L=22\pi$.
Figure~\ref{fig:baseline} shows the bifurcation diagram computed with ideal
(a,b: $\tau=0$) and realistic (c: $\tau=1$) boundary conditions. Bifurcation
points from the trivial solution (and secondary bifurcations from the primary
branches) are marked in the figure. The primary bifurcation points occur at
$\mu=\mu_{\smallpf}$ (\ref{eq:mupitchfork}), with the bifurcating branch being
locally proportional to $\sqrt{\mu-\mu_{\smallpf}}\sin(n\pi/L)$. With
$L=22\pi$, the solution with $n=11$ bifurcates precisely from $\mu=0$ and
corresponds to a pattern with 11~vortices in the half domain. The next four
bifurcating branches for positive $\mu$ are also shown in the figure. These
bifurcate at $\mu=0.0301$, $0.0361$, $0.1093$ and $0.1574$ and correspond to
$n=10$, 12, 9, and 13 vortices respectively. Note that these solutions are
invariant under reflections in the midpoint ($x=0$), and are spatially periodic
with period~$2L/n$.

Figure~\ref{fig:baseline}(c) shows a few of the many branches that exist for
realistic boundary conditions ($\tau=1$), in the same parameter range as
figure~\ref{fig:baseline}(b). These were obtained by taking all the points on
the branches in figure~\ref{fig:baseline}(a) with $\mu=2$ and continuing these
to $\tau=1$, and then continuing in~$\mu$ once more. Note that all of the
pitchfork bifurcations have been destroyed, and have been replaced by a series
of saddle-node bifurcations. The fundamental pitchfork bifurcation at $\mu=0$
has been replaced by a smooth transition, though the remnant of the pitchfork
shape can clearly be seen, and there is a sharp rise in amplitude close to
$\mu=0$ as predicted by the linear theory of the preceding section. Nonlinear
solution profiles on this fundamental branch
(figure~\ref{fig:ideal_real_examples}b) are qualitatively similar to those
with ideal boundary condition (figure~\ref{fig:ideal_real_examples}a), and so
might be said to correspond to $n=11$ vortices.

\begin{figure}
\begin{center}
\mbox{\psfig{file=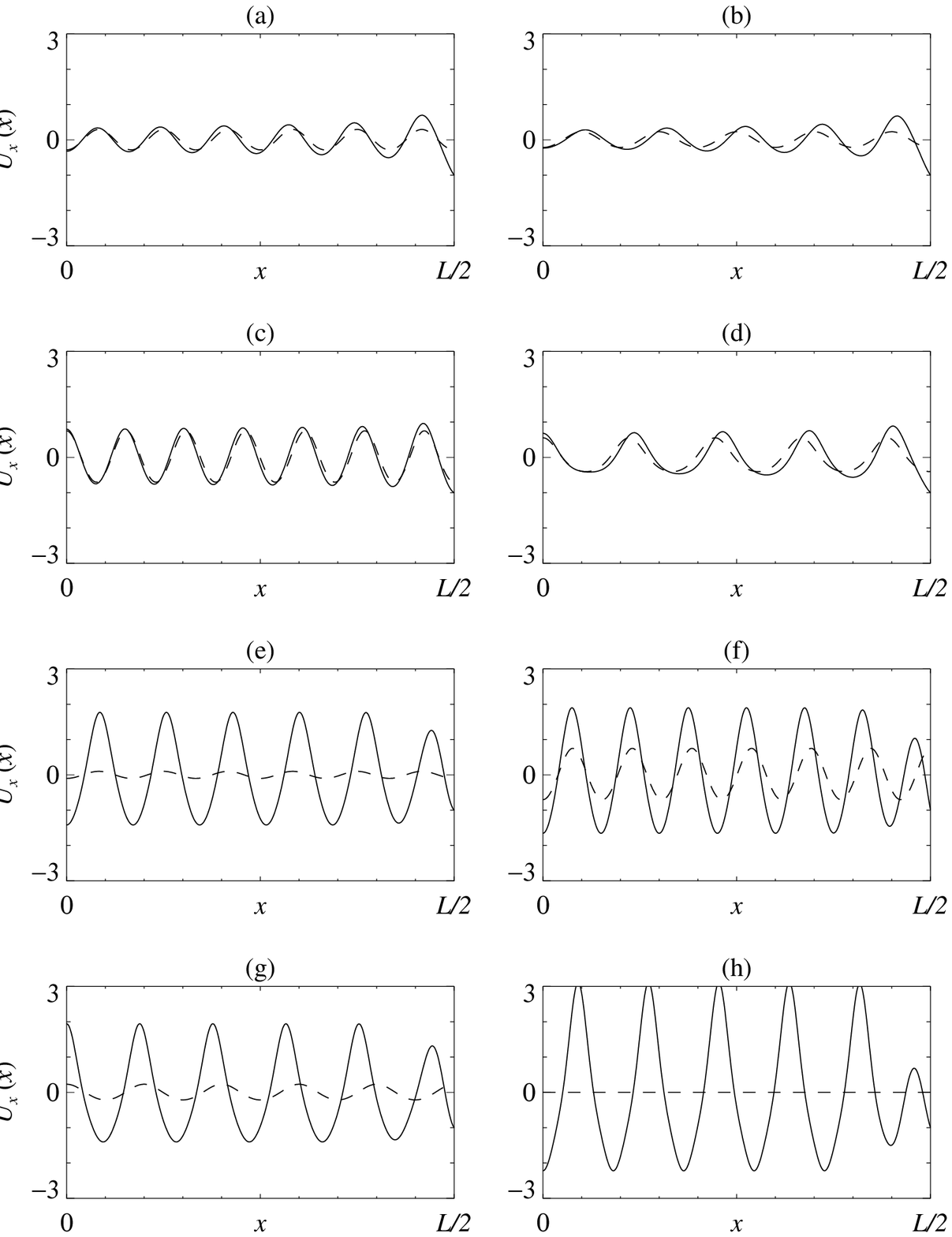,width=0.9\hsize}}
\end{center}
\caption{Examples of anomalous solutions at the labelled saddle-node
bifurcation points from figure~\ref{fig:baseline}(c). (g,h) are outside the
range shown in figure~\ref{fig:baseline}. Solid lines depict the solution at
the saddle node bifurcation with $\tau=1$, and dashed lines show the solution
that has been continued to $\tau\rightarrow0$, ending up at one of the
bifurcation points in figure~\ref{fig:baseline}(a).
 (a)~$\mu=0.1367\rightarrow0.0901$;
 (b)~$\mu=0.1498\rightarrow0.0770$;
 (c)~$\mu=0.5193\rightarrow0.4452$;
 (d)~$\mu=0.5581\rightarrow0.3787$;
 (e)~$\mu=1.6794\rightarrow0.0425$;
 (f)~$\mu=1.8227\rightarrow0.4452$;
 (g)~$\mu=2.1318\rightarrow0.0770$;
 (h)~$\mu=4.6220\rightarrow0.0000$.}
\label{fig:lp_examples}
\end{figure}

\begin{figure}
\begin{center}
\mbox{\psfig{file=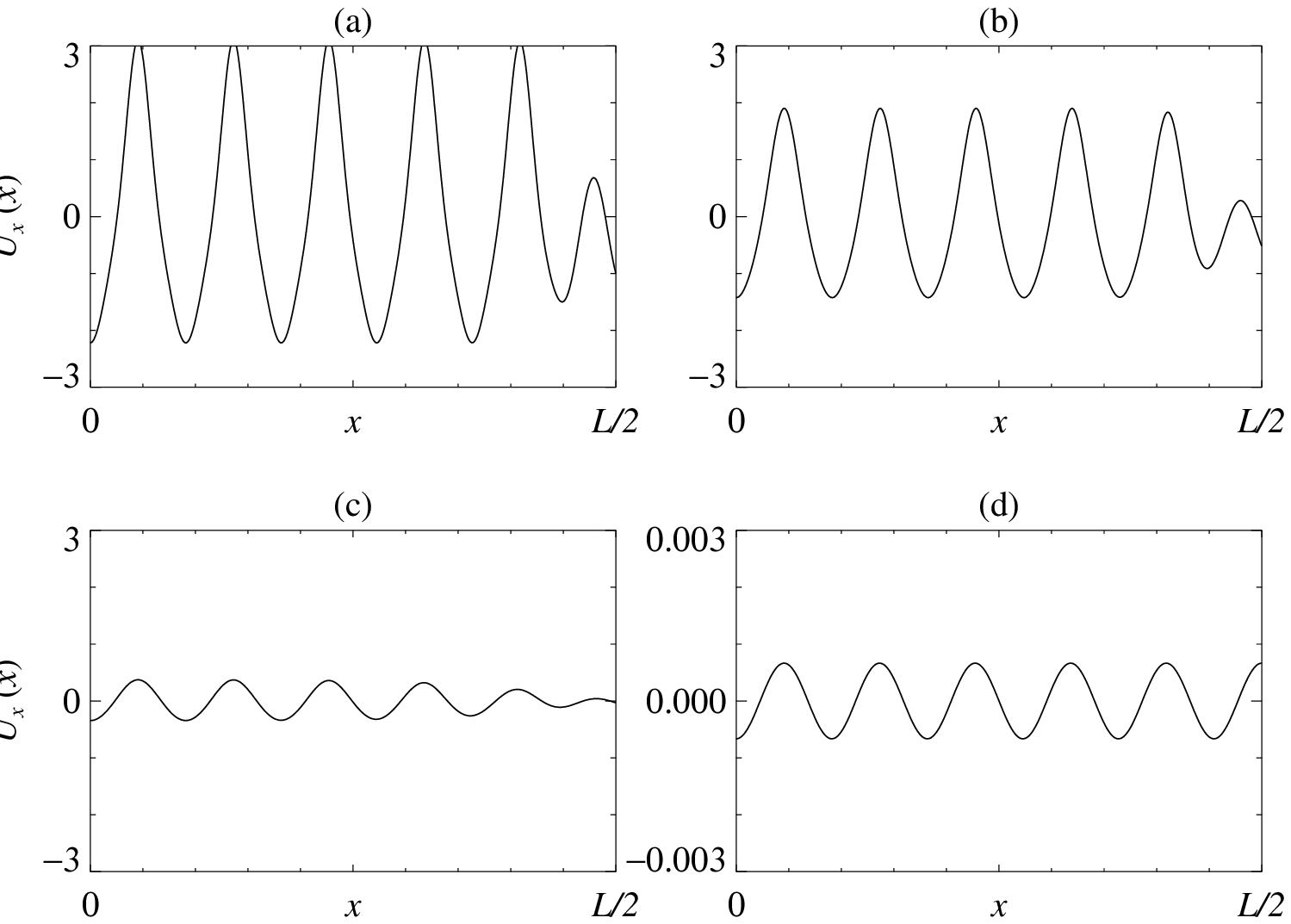,width=0.9\hsize}}
\end{center}
\caption{Continuation of the saddle-node bifurcation:
 (a)~$(\mu,\tau)=(4.622,1.000)$,
 (b)~$(\mu,\tau)=(2.000,0.597)$,
 (c)~$(\mu,\tau)=(0.100,0.076)$,
 (d)~$(\mu,\tau)\approx(0.000,0.000)$.}
\label{fig:lp_to_zero}
\end{figure}

This now brings us to the question of the supposed disconnected part of the
unfolded pitchfork. Examples of solutions at the labelled saddle-node
bifurcation points are shown in figure~\ref{fig:lp_examples}(a--h), with the
solution $U_x(x)$ at the saddle-node bifurcation points drawn as solid lines.
These saddle-node bifurcations were then continued in $(\mu,\tau)$ back to
$\tau=0$ to discover where they originate. Most connect to secondary
bifurcations in figure~\ref{fig:baseline}(a) and therefore represent unstable
solutions (and are shown as dashed lines in figure~\ref{fig:lp_examples}).
However, the saddle-node bifurcation on the stable anomalous 11~vortex branch,
shown in figure~\ref{fig:lp_examples}(e) is found to create a stable branch.
(The stability of a typical solution on this branch was checked by solving, in
addition to the ODEs  (\ref{steady}), the linear variational equations 
governing by a temporal eigenmode with eigenvalue~$s$. AUTO was then used to
continue solutions in~$s$ to establish that there are no nontrivial solutions
for~$s>0$.) Unexpected behaviour was found for this branch upon varying the
homotopy parameter~$\tau$. One might imagine that under homotopy to $\tau=0$
this saddle-node bifurcation point should approach the fundamental pitchfork
bifurcation at $\mu=0$. This is not the case for these parameter values: when
$\tau$ is reduced from 1 to 0, the saddle-node bifurcation itself undergoes a
pair of fold bifurcations, and ends up (at $\tau=0$) in the unfolding of the
$n=12$ bifurcation point at $\mu=0.0425$ (the dashed profile in
figure~\ref{fig:lp_examples}(e) corresponds to a 12~vortex pattern).

Alternatively, one could try following the saddle-node bifurcation point that
occurs as one unfolds the pitchfork at $\mu=0$ under infinitesimal increase of
$\tau$ from zero. When this is done, the saddle-node bifurcation can be
continued up to $(\mu,\tau)=(4.6220,1)$ -- see figure~\ref{fig:lp_examples}(h)
and figure~\ref{fig:lp_to_zero}. 

The details of which saddle-node bifurcation (with $\tau=1$) connects to which
pitchfork bifurcation (with $\tau=0$) was found to depend sensitively on the
value of~$L$. For example, with $L=88\pi$, the saddle-node bifurcation on the
44~vortex anomalous branch does continue down to the primary pitchfork to
44~vortices at $\mu=0$, though in this case the anomalous branch is not stable.
The details of how the branches connect also depends on the particular choice
how the ideal and realistic boundary conditions are combined via homotopy. For
example replacing the final `$+$' in (\ref{eq:BCboth}) with a `$-$' lead to
significantly different results.

\section{Conclusion}

The apparent contradiction described in the introduction is resolved, at least
in the context of the Swift--Hohenberg model considered here. The onset of
Taylor vortices is not a weakly broken pitchfork bifurcation, owing to strong
inhomogeneous boundary forcing. Anomalous modes stay bounded away from the
critical value of the bifurcation parameter as they must overcome the strong
preference set by the boundary. When the amplitude of the pattern is measured
far away from the boundaries, the pattern appears to set in sharply, in half a
pitchfork bifurcation, as the decaying wall mode penetrates the bulk of the
domain. The parameter value at which pattern, as measured in the centre of a
large domain, become significantly different from zero is the same as the value
predicted assuming idealised boundary conditions, because the requirements for
both situations are the same: steady ($s=0$) and zero spatial growth rate
($\sigma=0$). Moreover we have shown that for a long but finite domain of
length~$L$, that the parameter value corresponding to this large central growth
in pattern occurs according to linear theory at a value that is within
$O(L^{-2})$ of the idealised pitchfork.

The ideas were developed for the model equation, but they apply equally well to
the Taylor--Couette case, and resolve the difficulties raised by Benjamin and
Mullin~\cite{refB128}.

We have also observed that the saddle-node bifurcation on the anomalous branch
does not necessarily connect to the primary bifurcation as boundary conditions
vary from real to ideal. The specific results described here apply only to the
Swift--Hohenberg model, though the general conclusion that we can make is that,
under small perturbations from the ideal boundary conditions, we expect the
pitchfork to be perturbed in the generic way, as in
figure~\ref{fig:brokenpitchfork}(b). However, going all the way to $\tau=1$ is
not a small perturbation, and in general nothing can be said about whether the
the saddle-node bifurcation created in the unfolding of the primary pitchfork
is the same (or not the same) as the saddle-node at the end of the stable
anomalous branch, or even if there is an anomalous branch that is stable.

This approach yields results that are applicable to other pattern formation
problems (for instance, Rayleigh--B\'enard convection). Earlier work on
convection~\cite{refD56,DaHoSk:03} has focused on weak forcing at the side
walls, primarily using Swift--Hohenberg theory. We have shown here how the
ideas can be extended to strong forcing.

An interesting aspect of our work has been to link the mode selection problem
to the existence of a spatial Hamiltonian Hopf bifurcation for the infinite
length problem. This gives the possibility of the existence of branches of
spatially periodic solutions beyond the critical parameter value for the onset
of rolls, whose period for small amplitude is that given by the wave number of
the neutral mode of the temporal problem. The normal form of the Hamiltonian
Hopf bifurcation (even for reversible systems) is completely integrable up to
any order (see e.g., \cite{IoPe:93}) and instead of a unique spatially periodic
solution, there is a one parameter band of spatially periodic solutions, whose
envelope grows as the square root of the bifurcation parameter~$\mu$. There
also exists a two-parameter family of spatially quasi-periodic solutions whose
existence is bounded by the periodic solutions and homoclinic connections to
them. However, in the reversible case, not all these solutions will necessarily
exist in a full unfolding of the normal form that breaks its Hamiltonian
structure. We note from our numerical results for the nonlinear problem with
realistic boundary conditions and fixed~$L$, that the main branch seems to
develop a connection from the boundary to a pure periodic state in the middle
of the domain. It is not clear {\em a priori} why this solution and not others
are selected from the unfolding of the Hamiltonian Hopf bifurcation.

Finally, we mention another outstanding issue in the problem of the onset of
Taylor vortices~\cite{refA70}: the timescale for the onset and decay of the
pattern are different. In particular, these authors computed steady solutions
at Reynold's numbers just above and just below critical, and in each case
altered the Reynold's number to an intermediate value and examined the
transient. In the case of onset, the pattern invaded the bulk as a front
travelling in from the boundary, while in the case of decay, there was uniform
decay throughout bulk. The timescales for these two processes were different,
and it may be possible to explain this using a Swift--Hohenberg based model, as
considered here.

 \begin{ack}
AMR is grateful for support from the EPSRC while this work was carried out. We
are very grateful to Tom Mullin for many coments and advice, and we thank Edgar
Knobloch for useful discussions.
 \end{ack}


\end{document}